\documentclass[sigconf,authorversion]{acmart}

\usepackage{placeins} 

\usepackage{adjustbox}

\usepackage{stfloats}

\usepackage[colorinlistoftodos,prependcaption]{todonotes} 
\presetkeys%
    {todonotes}%
    {inline,backgroundcolor=orange}{}
    
\usepackage{fontawesome}
\usepackage[nopostdot,acronym,shortcuts,nonumberlist]{glossaries}
\newacronym{(S+N)/NR}{(S+N)/NR}{signal-plus-noise-to-noise ratio}
\newacronym{ADC}{ADC}{analog-to-digital converter}
\newacronym{BCI}{BCI}{bulk current injection}
\newacronym{CAN}{CAN}{control area network}
\newacronym{CRT}{CRT}{cathode ray tube}
\newacronym{CSMACD}{CSMA/CD}{carrier sense multiple access with collision detection}
\newacronym{CoE}{CoE}{CAN over Ethernet}
\newacronym{DAC}{DAC}{digital-to-analog converter}
\newacronym{DoS}{DoS}{Denial of Service}
\newacronym{EMR}{EMR}{electromagnetic radiation}
\newacronym{FCS}{FCS}{frame check sum}
\newacronym{FEC}{FEC}{forward error correction}
\newacronym{FPGA}{FPGA}{field-programmable gate array}
\newacronym{ICMP}{ICMP}{Internet control message protocol}
\newacronym{IEC}{IEC}{International Electrotechnical Commission}
\newacronym{MAC}{MAC}{Medium Access Control}
\newacronym{NIC}{NIC}{network interface card}
\newacronym{NSA}{NSA}{National Security Agency}
\newacronym{PA}{PA}{power amplifier}
\newacronym{PC}{PC}{personal computer}
\newacronym{RF}{RF}{radio frequency}
\newacronym{SDR}{SDR}{software-defined radio}
\newacronym{SFD}{SFD}{start frame delimiter}
\newacronym{SNR}{SNR}{signal-to-noise ratio}
\newacronym{USRP}{USRP}{Universal Software Radio Peripheral}
\newacronym{SSP}{SSP}{Secure Simple Pairing}
\newacronym{USB}{USB}{Universal Serial Bus}
\newacronym{TLV}{TLV}{Type Length Value}
\newacronym{L2CAP}{L2CAP}{Logical Link Control and Adaptation Protocol}
\newacronym{AOP}{AOP}{Always-On Processor}
\newacronym{ACL}{ACL}{Asynchronous Connec\-tion-Less}
\newacronym{BLE}{BLE}{Bluetooth Low Energy}
\newacronym{BR}{BR}{Basic Rate}
\newacronym{CID}{CID}{Channel ID}
\newacronym{PDU}{PDU}{Protocol Data Unit}
\newacronym{B-Frame}{B-Frame}{Basic Information Frame}
\newacronym{C-Frame}{C-Frame}{Control Frame}
\newacronym{PSM}{PSM}{Protocol/Service Multiplexer}
\newacronym{HCI}{HCI}{Host Controller Interface}
\newacronym{EDR}{EDR}{Enhanced Data Rate}
\newacronym{SCO}{SCO}{Synchronous Connection-Oriented}
\newacronym{HSP}{HSP}{Headset Profile}
\newacronym{ATT}{ATT}{Attribute Protocol}
\newacronym{TCP}{TCP}{Transmission Control Protocol}
\newacronym{SoC}{SoC}{System-on-a-Chip}
\newacronym{IoT}{IoT}{Internet of Things}
\newacronym{GPS}{GPS}{Global Positioning System}
\newacronym{CERT}{CERT}{Computer Emergency Response Team}
\newacronym{API}{API}{Application Programming Interface}
\newacronym{MITM}{MITM}{Machine-in-the-Middle}
\newacronym{TLS}{TLS}{Transport Layer Security}
\newacronym{OTA}{OTA}{Over-the-Air}
\newacronym{XSS}{XSS}{Cross-Site Scripting}
\newacronym{SDK}{SDK}{Software Development Kit}
\newacronym{MQTT}{MQTT}{Message Queuing Telemetry Transport}
\newacronym{CVE}{CVE}{Common Vulnerabilities and Exposures}
\newacronym{ECDH}{ECDH}{Elliptic-curve Diffie–Hellman}
\newacronym{AEAD}{AEAD}{Authenticated Encryption with Associated Data}
\newacronym{AES}{AES}{Advanced Encryption Standard}
\newacronym{TOR}{TOR}{The Onion Router}
\newacronym{GDPR}{GDPR}{EU General Data Protection Regulation}
\newacronym{HSTS}{HSTS}{HTTP Strict Transport Security}
\newacronym{AWS}{AWS}{Amazon Web Services}
\newacronym{LTK}{LTK}{Long Term Key}
\newacronym{LK}{LK}{Link Key}
\newacronym{RCE}{RCE}{Remote Code Execution}
\newacronym{SIV}{SIV}{Synthetic Initialization Vector}
\newacronym{PoC}{PoC}{Proof of Concept}
\newacronym{XPC}{XPC}{Cross-Process Communication}
\newacronym{MFi}{MFi}{Made for iPhone/iPad/iPod}
\newacronym{ACI}{ACI}{Apple Controller Interface}
\newacronym{ECB}{ECB}{Electronic Codebook}
\newacronym{UART}{UART}{Universal Asynchronous Receiver-Transmitter}

\newacronym{KDF}{KDF}{Key Derivation Function}
\newacronym{PRF}{PRF}{Pseudo-Random Function}

\tikzstyle{line} = [draw, -latex']
\usetikzlibrary{shapes,arrows,calc,positioning,matrix}
\usetikzlibrary{positioning}
\tikzset{
    >=triangle 45
}

\definecolor{darkred}{rgb}{0.831, 0, 0.063}

\usepackage[binary-units]{siunitx}
\usepackage{tabularx}

\definecolor{sorange}{rgb}{0.95, 0.57, 0}
\colorlet{orange}{sorange}

\usepackage{listings}
\lstdefinelanguage{ASM}{
    morekeywords={b, ble, blt, bne, bx, bl, ldr, str, push, pop, mov, add, sub},
    keywordstyle=\color{blue},
    sensitive=false, 
    morecomment=[l]{//}, 
    morecomment=[s]{/*}{*/}, 
    morestring=[b]", 
} %
\lstdefinelanguage{none}{
  identifierstyle=
}

\usepackage{subcaption}

\usepackage{xspace}

\usepackage{bytefield}
\usepackage{xcolor,colortbl}
	
\acmYear{2020}\copyrightyear{2020}
\setcopyright{rightsretained}
\acmConference[WiSec '20]{13th ACM Conference on Security and Privacy in Wireless and Mobile Networks}{July 8--10, 2020}{Linz (Virtual Event), Austria}
\acmBooktitle{13th ACM Conference on Security and Privacy in Wireless and Mobile Networks (WiSec '20), July 8--10, 2020, Linz (Virtual Event), Austria}
\acmPrice{15.00}
\acmDOI{10.1145/3395351.3401697}
\acmISBN{978-1-4503-8006-5/20/07}

\acmSubmissionID{2}

\begin{document}

\title[InternalBlue on macOS]{DEMO: Attaching InternalBlue to the Proprietary\\ macOS IOBluetooth Framework}


\author{Davide Toldo}
\affiliation{%
  \institution{Secure Mobile Networking Lab}
  \institution{TU Darmstadt, Germany}
  \streetaddress{Pankratiusstraße 2}
  \postcode{64289}
}
\email{dtoldo@seemoo.de}

\author{Jiska Classen}
\affiliation{%
  \institution{Secure Mobile Networking Lab}
  \institution{TU Darmstadt, Germany}
  \streetaddress{Pankratiusstraße 2}
  \postcode{64289}
}
\email{jclassen@seemoo.de}

\author{Matthias Hollick}
\affiliation{%
  \institution{Secure Mobile Networking Lab}
  \institution{TU Darmstadt, Germany}
  \streetaddress{Pankratiusstraße 2}
  \postcode{64289}
}
\email{mhollick@seemoo.de}

\renewcommand{\shortauthors}{Toldo et al.}


\begin{abstract}
In this demo, we provide an overview of the \emph{macOS} Bluetooth stack internals and gain access to undocumented low-level interfaces.
We leverage this knowledge to add \emph{macOS} support to the \emph{InternalBlue} firmware modification
and wireless experimentation framework.
\end{abstract}



\maketitle

\section{Introduction}
The \emph{macOS} Bluetooth stack is an interesting research target, since all \emph{iMacs}
and \emph{MacBooks} exclusively use \emph{Broadcom} Bluetooth chips. These chips
allow unsigned temporary firmware patches via \emph{InternalBlue}~\cite{mantz2019internalblue}.
Integrating \emph{macOS} into \emph{InternalBlue} enables full access 
to the Bluetooth chips in hundreds of millions of devices.

Officially, Bluetooth access on \emph{macOS} is supported by the
\path{CoreBluetooth} and \path{IOBluetooth} frameworks. However, these frameworks are very restricted.
Firmware modification requires access to the \ac{HCI}, and sending arbitrary data
over-the-air requires \ac{ACL} injection. 
%
We reverse-engineer the \emph{macOS} Bluetooth stack to understand \ac{HCI} and \ac{ACL} in \autoref{sec:stack}. Based on the reverse-engineering results, we develop custom hooks that could also
be extended for other applications in \autoref{sec:hook}.
We conclude the results of the \emph{macOS} \emph{IntenalBlue} integration in \autoref{sec:conclusion}.

\newpage 

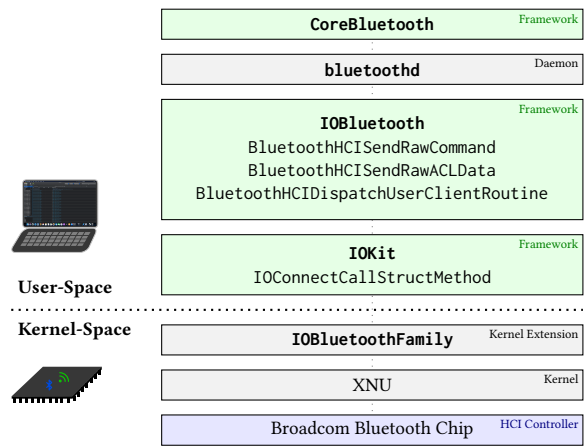
\begin{figure}[!b]
\center

	\begin{center}
	\begin{tikzpicture}[minimum height=0.55cm, scale=0.8, every node/.style={scale=0.8}, node distance=0.7cm]
	\tikzset{w/.style={fill=white, align=center}}
	\tikzset{block/.style={minimum width = 7cm, minimum height = 0.5cm, draw=black, fill=white, align=center}}
	\tikzset{framework/.style={block, fill=green!10}}
	
    \draw[-,black,dotted,thick] (-9.5,2.25) -- (0,2.25);
    \node[align=right,right] (top) at (-9.5, 2.6) {\textbf{User-Space}};
    \node[align=right,right] (down) at (-9.5, 1.9) {\textbf{Kernel-Space}};
    
    
    \draw[-,gray,dotted] (-3.5,0.25) -- (-3.5,7);
	
	 \node[block, name=bb, fill=blue!10] at (-3.5,0.25) {Broadcom Bluetooth Chip};
	 \node[block, name=x, fill=gray!10] at (-3.5,1) {XNU};
	 \node[block, name=if, fill=gray!10] at (-3.5,1.75) {\texttt{\textbf{IOBluetoothFamily}}};
	 \node[framework, name=ik, minimum height = 1cm] at (-3.5,3) {\texttt{\textbf{IOKit}}\\\texttt{IOConnectCallStructMethod}};
	 \node[framework, name=ib, minimum height = 2cm] at (-3.5,4.75) {\texttt{\textbf{IOBluetooth}}\\\texttt{BluetoothHCISendRawCommand}\\\texttt{BluetoothHCISendRawACLData}\\\texttt{BluetoothHCIDispatchUserClientRoutine}};
	 \node[block, name=bd, fill=gray!10] at (-3.5,6.25) {\texttt{\textbf{bluetoothd}}};
	 \node[framework, name=cb] at (-3.5,7) {\texttt{\textbf{CoreBluetooth}}};
	 
	 \node[below left=-0.1cm and -0.025cm] at (cb.north east) {\scriptsize \textcolor{green!50!black}{Framework}};
	 \node[below left=-0.1cm and -0.025cm] at (bd.north east) {\scriptsize Daemon};
	 \node[below left=-0.1cm and -0.025cm] at (ib.north east) {\scriptsize \textcolor{green!50!black}{Framework}};
	 \node[below left=-0.1cm and -0.025cm] at (ik.north east) {\scriptsize \textcolor{green!50!black}{Framework}};
	 \node[below left=-0.1cm and -0.025cm] at (if.north east) {\scriptsize Kernel Extension};
	 \node[below left=-0.1cm and -0.025cm] at (x.north east) {\scriptsize Kernel};
	 \node[below left=-0.1cm and -0.025cm] at (bb.north east) {\scriptsize \textcolor{blue!50!black}{HCI Controller}};
	 
    \node[inner sep=0pt,align=right,right] (laptop) at (-9.5,3.8)
    {\includegraphics[width=1.5cm]{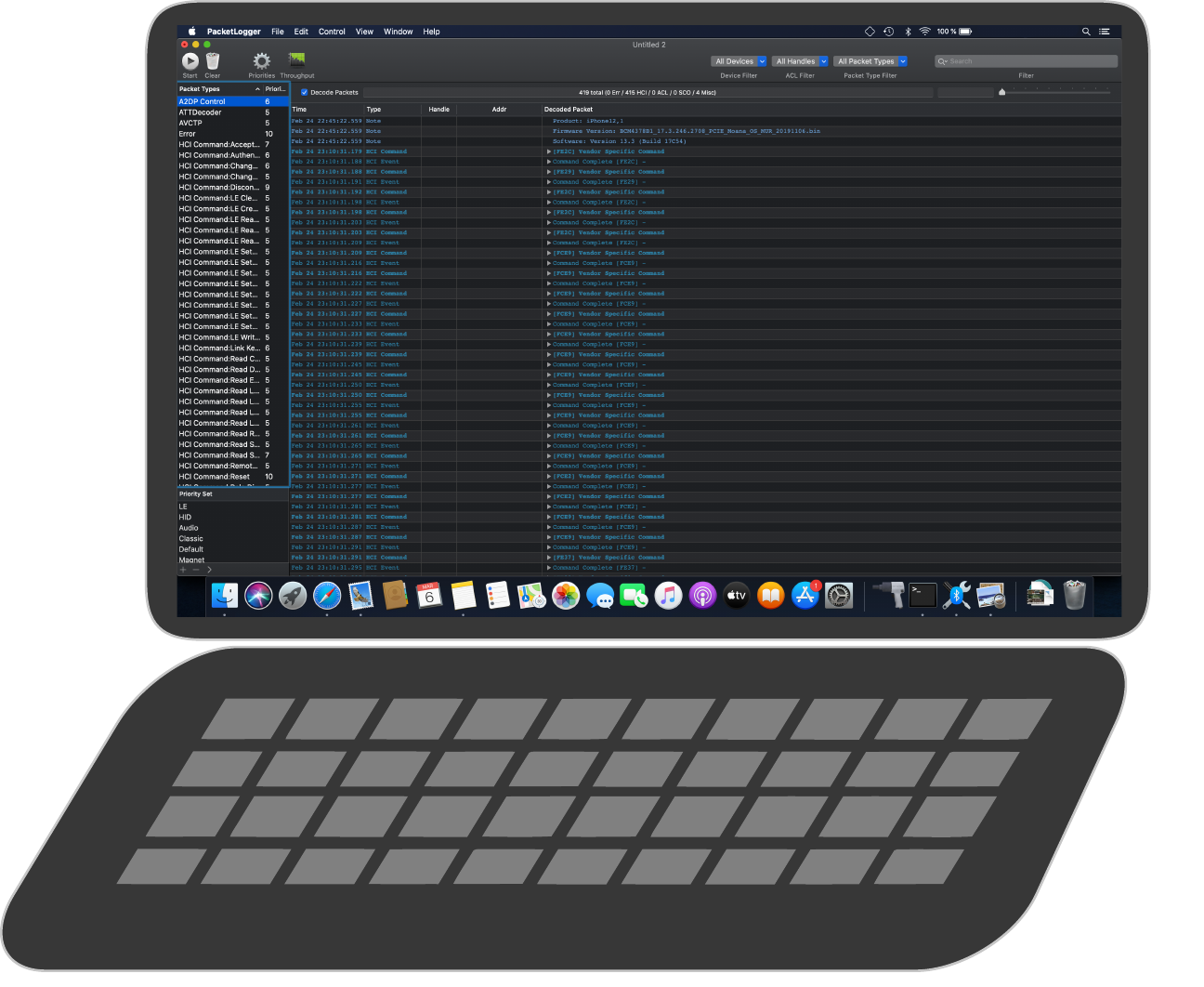}};   
    \node[inner sep=0pt,align=right,right] (btchip) at (-9.5,1)
    {\includegraphics[width=1.5cm]{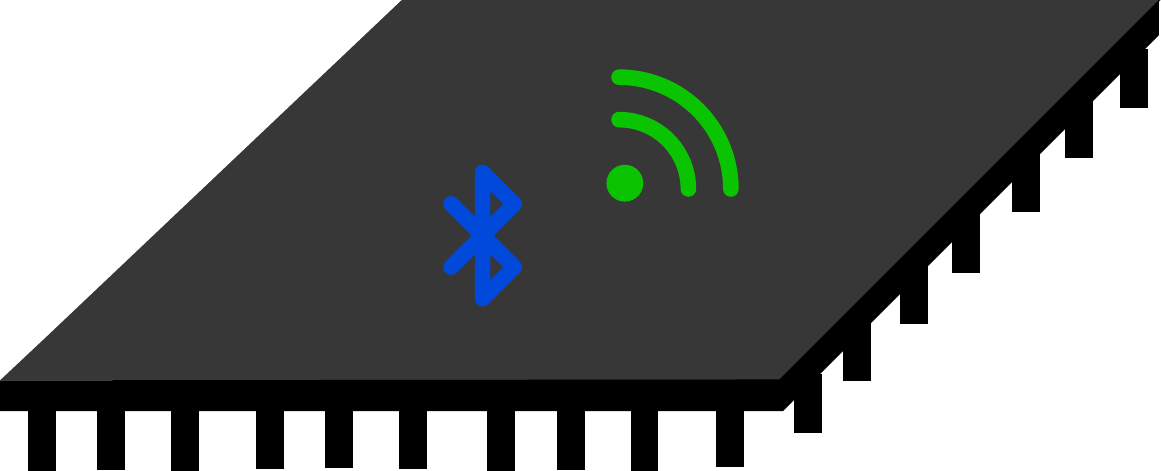}};

	\end{tikzpicture}
	\end{center}
	
\caption{\emph{Apple's}  \emph{macOS} Bluetooth stack.}
\label{fig:stacksimple}
\end{figure}

\section{Bluetooth Stack Overview}
\label{sec:stack}

An overview of the \emph{macOS} Bluetooth stack is shown in \autoref{fig:stacksimple}.
User-space applications do not interact directly with the chip, \emph{mac\-OS} restricts  communication to the \texttt{IOBluetoothFamily.kext} driver, running in kernel-space.
The official \emph{macOS} Bluetooth API does not allow sending \ac{HCI}, \ac{ACL}, or \ac{SCO} packets.
However, \path{CoreBluetooth} and selected publicly documented classes and functions of \path{IOBluetooth} offer application developers high-level access to a few very basic functions, i.e., retrieving the Bluetooth address~\cite{iobt,corebt}.
Playing music via Bluetooth headphones is abstracted further, as the application developer only needs to be aware of music playback but not the
specific output method. Thus, audio functions can be accessed via \path{AVAudioPlayer}, and then, \mbox{\emph{macOS}} decides if the music is sent to the internal
speakers or an external Bluetooth peripheral depending on the audio settings selected by the user. In case Bluetooth headphones are connected and selected,
the music is forwarded to \path{bluetoothaudiod}. Since it is a separate daemon, it forwards the audio again to \path{bluetoothd} via \ac{XPC}.

Nonetheless, the various Bluetooth frameworks need to access functions within \path{IOBluetooth}, specifically \path{BluetoothHCISendRawCommand}
for sending \ac{HCI} commands to the Bluetooth chip and \path{BluetoothHCISendRawACLData} for transmitting \ac{ACL} data.
\path{CoreBluetooth} only accesses these functions indirectly via \path{bluetoothd}, which in turn accesses \path{IOBluetooth}.

Note that the function to send \ac{HCI} commands was reverse-engineered and documented before~\cite{hcitool}.
However, the existing project only supported selected commands documented in the Bluetooth specification and had no
external interface, while \emph{InternalBlue} support requires vendor-specific commands. Moreover, we are the
first to reverse-engineer \ac{ACL} on \emph{macOS}.


\begin{figure*}[!t]
\begin{minipage}{0.88\textwidth}
\begin{lstlisting}[language=C, caption=Reverse-engineered function names., label=lst:functions]
int BluetoothHCISendRawCommand(uint32_t request, void *commandData, size_t commmandSize);
int BluetoothHCISendRawACLData(void *commandData, size_t commandSize, uint32_t handle, uint32_t request);
\end{lstlisting}
\end{minipage}
\end{figure*}

\begin{figure*}[!t]
	\footnotesize
	\fontfamily{\ttdefault}\selectfont
	\center

    \begin{subfigure}[b]{\textwidth}
		\center
		\begin{tabular}{ |l|l|l|l|l| } 
        \hline
	    \rowcolor{gray!10}
        \textcolor{red}{Apr 22 23:40:49:668} & \textcolor{red}{Error} & ~ & \hspace{1em} & \hspace{0.7em} \textcolor{red}{ACLPacketToHw No Device Handle 0x172} \\
        \hline
        \textcolor{purple!70!blue}{Apr 22 23:40:49:668} & \textcolor{purple!70!blue}{LEAS Send} & \textcolor{purple!70!blue}{0x0172} & ~ & \textcolor{gray}{$\blacktriangleright$} \textcolor{purple!70!blue}{Data [Handle: 0x0172, Packet Boundary Flags: 0x3, Length: 0x0010 (16)]} \\
        \hline
	    \rowcolor{gray!10}
        \textcolor{red}{Apr 22 23:40:49:668} & \textcolor{red}{Error} & ~ & ~ & \hspace{0.7em} \textcolor{red}{Above ACL Packet not sent Handle 0x172} \\
        \hline
	    \end{tabular}
		\caption{ACL with wrong function parameters.}
		\label{fig:aclerror}
    \end{subfigure}    
    
    \begin{subfigure}[b]{\textwidth}

		\center
		\begin{tabular}{ |l|l|l|l|l| } 
        \hline
        \textcolor{purple!70!blue}{Apr 22 23:44:30.514} & \textcolor{purple!70!blue}{LEAS Send} & \textcolor{purple!70!blue}{0x000B} & \hspace{1em} & \textcolor{gray}{$\blacktriangleright$} \textcolor{purple!70!blue}{Data [Handle: 0x000B, Packet Boundary Flags: 0x3, Length: 0x0010 (16)]} \\
        \hline
	    \rowcolor{gray!10}
        \textcolor{blue}{Apr 22 23:44:31.006} & \textcolor{blue}{HCI Event} & \textcolor{blue}{0x000B} & ~ & \textcolor{gray}{$\blacktriangleright$} \textcolor{blue}{Number of Completed Packets - Handle: 0x000B - Packets: 0x0001} \\
        \hline
	    \end{tabular}
		\caption{Successful ACL transmission.}
		\label{fig:aclsuccess}
    \end{subfigure}   
\caption{ACL method calls captured with \emph{PacketLogger}.} 
\label{fig:packetlogger}

\end{figure*}

\ac{HCI} and \ac{ACL} are slightly different in their functionality. For example, \ac{HCI} supports configuring the Bluetooth
chip. Most \ac{HCI} commands are not connection-related. In contrast, \ac{ACL} is used for data
transmission within an active connection and, thus, always requires a connection handle. However, communication with the Bluetooth chip's interface
is very similar for both of them. Thus, both use the more generic private \path{BluetoothHCIDispatchUserClientRoutine}, which passes them to the
\path{IOKit} user-space framework~\cite{iokit}. The corresponding function is called \path{IOConnectCallStructMethod} and finally forwards the Bluetooth
packet to the \texttt{IOBluetoothFamily} kernel-space driver using a Mach port. This driver supports various means of transportation to the chip: USB, UART, and PCIe.

The \ac{HCI} and \ac{ACL} methods within \path{IOBluetooth} are not callable by external binaries because they are not declared in the
\path{IOBluetooth} headers. By declaring them in a header file of an \emph{Objective-C} project and importing the framework, they become callable. While we chose the
methods within \path{IOBluetooth} to support \emph{InternalBlue} on \emph{macOS}, it would also be possible to instead hook into and import \path{IOKit} and
use the \path{IOConnectCallStructMethod}, which is even deeper in the stack.
With this declaration, any user on \emph{macOS} is able to execute a binary that calls these functions---no privileged access
is required to modify the firmware on the Bluetooth chip.

An alternate approach is to communicate with \path{bluetoothd} via \ac{XPC}~\cite{bleno}. However, our approach bypasses \path{bluetoothd} and directly communicates with the chip---no capability checks on the calling process are performed.

Overall, \path{bluetoothd} has more of an administrative role on \emph{macOS}. In contrast, \path{bluetoothd} on \emph{iOS} is located much deeper within the stack~\cite{magicpairing}.


\section{Reverse-Engineering Techniques}
\label{sec:hook}
We used various reverse-engineering tools and debugging methods to analyze the \emph{macOS} Bluetooth stack.

Initially, we analyzed \emph{macOS} binaries using \emph{Hopper v4} and \emph{Ghidra}. Most of these binaries
were not stripped and still contained most function names, enabling full-text searches for \texttt{`ACL'} and
\texttt{`HCI'}. As \path{bluetoothd} excessively calls the private \path{IOBluetooth} framework, this provided
us with many insights.

When accessing functions like \path{BluetoothHCISendRawCommand} within a project, they need to be declared in
an \emph{Objective-C} header file and \path{IOBluetooth} has to be imported. By importing the framework, the binary
is linked against it---which also includes undocumented methods.
However, to call functions within the \mbox{\path{IOBluetooth}} framework, not only their names but also their
precise arguments and types are required. 
There are two methods to reverse-engineer these. The most commonly known is runtime analysis with a
debugger like \path{lldb}. However, we chose another option. \emph{Apple} provides a Bluetooth \emph{PacketLogger}
in their \emph{Additional Tools for Xcode}.

By trying different data types and values for the arguments of \path{BluetoothHCISendRawCommand} and \path{BluetoothHCISendRawACLData} and
simultaneously checking the logs in \emph{PacketLogger}, we were able to determine the purposes and data types of each variable and reconstructed the function signatures.
As shown in \autoref{lst:functions}, both functions have parameters for a request identifier, the data to be transmitted, and the total command size in bytes.
\ac{ACL} connections are always end-to-end with another device, which is identified by a handle to support multiple \ac{ACL} connections in parallel.
Thus, the \ac{ACL} function requires an additional handle parameter.

When guessing the correct function signatures, \emph{PacketLogger} provides immediate feedback as shown in \autoref{fig:packetlogger}.
For example, we initially swapped the \path{handle} and \path{request} identifier. Thus, \emph{PacketLogger} complains that there is
no connection with the handle \path{0x0172}. After swapping these parameters, \ac{ACL} data can be transmitted successfully.

\section{Conclusion}
\label{sec:conclusion}

We tested the resulting \emph{macOS} \emph{InternalBlue} port on a high variety of devices,
including a recent \emph{MacBook Pro 16'' Late 2019} on \emph{Catalina}, going back to an \emph{iMac Late 2009} on \emph{High Sierra}.
Thus, even though the \path{IOBluetooth} framework is undocumented,
\ac{HCI} and \ac{ACL} access stay the same across various \emph{macOS} versions.
Moreover, using \path{IOBluetooth} operates independently from the underlying transport mode, which can be USB, UART, or PCIe.
With the approach described in this demo, \emph{InternalBlue} works on all these chip variants.
This enables Bluetooth security research on various devices.

\section*{Demo Setup}
Our demonstration will consist of two parts: \emph{(1)} a minimal working example to hook \path{IOBluetooth} functions on \emph{macOS}, as well as \emph{(2)} a video
recording of the full \emph{InternalBlue} integration.

The minimal working example requires access to a \emph{macOS} computer running \emph{Mojave} or \emph{Catalina} and \emph{Xcode}.
This example provides the user with a command-line application that demonstrates how to call private \path{IOBluetooth} functions.
In contrast, while the \emph{InternalBlue} integration uses the same mechanism, is way more complex and harder to understand.
The code of this example is openly available and we provide detailed installation and usage instructions.

Since not everyone has access to a \emph{macOS} device or the time to compile a project in \emph{Xcode}, we will also upload
videos of the minimal working example and the full \emph{InternalBlue} integration.

\newpage
\begin{acks}
This work has been funded by the German Federal Ministry of Education and Research and the Hessen State Ministry for Higher Education, Research and the Arts within their joint support of the National Research Center for Applied Cybersecurity ATHENE.
\end{acks}

\bibliographystyle{ACM-Reference-Format}
\bibliography{bibfile}

\end{document}